\documentclass[12pt]{article}
\pdfoutput=1

\usepackage{paper2e}
\usepackage{mydefs2e}
\usepackage{xspace}
\usepackage{graphicx}

\renewcommand{\d}{\partial}

\begin{document}

\begin{titlepage}

\title{Minimal Gaugomaly Mediation}

\author{Yi Cai,\ \ Markus A. Luty}

\address{Physics Department, University of California Davis\\
Davis, California 95616}

\begin{abstract}
Mixed anomaly and gauge mediation (``gaugomaly'' mediation)
gives a natural solution to the SUSY flavor problem with
a conventional LSP dark matter candidate.
We present a minimal version of gaugomaly mediation where
the messenger masses arise directly from anomaly
mediation, automatically generating a messenger scale of order
$50\TeV$.
We also describe a simple relaxation mechanism that
gives rise to realistic $\mu$ and $B\mu$ terms.
$B$ is naturally dominated by the anomaly-mediated contribution
from top loops, so the $\mu$-$B\mu$ sector only depends on a single
new parameter.
In the minimal version of this scenario the full
SUSY spectrum is determined by two continuous
parameters (the anomaly- and gauge-mediated SUSY breaking masses)
and one discrete parameter (the number of messengers).
We show that these simple models can give realistic spectra
with viable dark matter.
\end{abstract}

\end{titlepage}

\section{Introduction}
\label{sec:intro}
Anomaly-mediated SUSY breaking (AMSB) is a natural solution to the flavor
problem \cite{AMSB} that may be naturally realized in string theory
\cite{stringsequester}.
The minimal model with only the MSSM in the visible sector
predicts negative slepton masses.
However, \Ref{PR} pointed out that if the visible
sector contains messenger fields whose masses arise from
AMSB, they automatically give messenger masses with $F/M \sim F_\varphi$,
where $F_\varphi \sim 50\TeV$ is the anomaly-mediated order parameter.
This means that the gauge- and anomaly-mediated contributions are
automatically of the same size, and
this kind of model easily give a realistic SUSY breaking spectrum.
We call this ``gaugomaly'' mediation, following \Ref{gaugomaly}.
An important difference from gauge mediation is that the gravitino
is naturally heavy (of order $F_\varphi$)
so the LSP is generally
a neutralino, and therefore a WIMP dark matter candidate.
Gaugomaly mediation is probably
the simplest framework for SUSY breaking that
solves the SUSY flavor problem and has a natural dark matter candidate.

It was pointed out in \Ref{NW} that messenger masses can arise from
holomorphic \Kahler terms of the form
$\int d^4\th\, \bar\Phi \Phi + \mbox{h.c}$.
If the dimensionless coefficient of such a term is order 1
this gives a messenger threshold at the scale $F_\varphi$.
Unfortunately slepton masses remain negative at the messenger scale
in this model.
Running from the scale $F_\varphi$ down to the weak scale can give
positive slepton masses from the enhanced gaugino contributions
to the scalar masses for a sufficiently large threshold correction,
but this requires a large number of messengers and additional
fine-tuning.
\Ref{HL} showed that a successful model can easily
be obtained with additional singlet fields.
If these have order 1 holomorphic \Kahler terms they naturally
get SUSY breaking VEVs set by the scale $F_\varphi$.
Coupling these fields to messengers then gives a messenger threshold
that can give an acceptable spectrum with positive slepton masses
at the messenger threshold.


In this paper, we expand and improve upon \Ref{HL} in several
ways.
First, we construct simple models that generate
$\mu$ and $B\mu$ terms of the right size at the messenger
threshold.
Second, we compute the SUSY spectrum 
and show that there are solutions with viable WIMP dark matter.
We also include a discussion of tadpole terms for singlet
fields that were neglected in \Ref{HL}
(but do not change the results),
and give complete formulas for the SUSY breaking masses
to all orders in $F/M^2$.

The models we present are technically natural,
but some superpotential couplings allowed by symmetries
are absent.
There are good reasons to think that
technical naturalness may well be all that we should
expect in supersymmetric theories.
For example, if superpotential couplings arise from the
VEVs are chiral superfields, then any
technically natural theory is the most general theory
invariant under a set of symmetries,
some of which are broken by the VEVs \cite{SeibergNR}.
The mechanisms at work in our models are simple and robust,
and it is likely that they work in models that include
all terms allowed by symmetries.
Focusing on simple technically natural models allows us to
illustrate the basic mechanism in the simplest possible setting.

The general picture that emerges from this work is the following.
In AMSB the scale $F_\varphi$ is a natural scale for singlets
and vector-like matter to get masses and/or VEVs.
Such additional fields do not ruin gauge coupling
unification, and are ubiquitous in string constructions.
One may therefore expect many such fields with masses of order 
$F_\varphi$.
Some of these fields act as messengers,
and the spectrum will interpolate
between an anomaly-mediated spectrum and a gauge-mediated
spectrum.
This is a rather simple and appealing picture that accords
well with general theoretical expectations.

It also gives a very predictive spectrum of SUSY breaking masses.
The $\mu$ and $B\mu$ terms are generated via a relaxation
mechanism, but quite generally the contribution to the $B$ term
from this sector is smaller than the anomaly-mediated contribution,
which comes from a top loop.
Therefore, the $\mu$-$B\mu$ sector effectively contributes
a single free parameter, namely the value of $\mu$.
The SUSY breaking spectrum is then specified by 
$F_\varphi$, a gauge mediated order parameter $F / M \sim F_\varphi$,
$\mu$, and the number of messengers $N$.
There is a weak dependence on the ratio
$x = F / M^2$ that is unimportant except near $x = 1$.
Once we fix the Higgs VEV to its experimental value, the
minimal model depends on only two continuous parameters, and the
number of messengers.
We find regions of this parameter space where the dark
matter relic abundance has the correct value,
and dark matter direct detection is below current bounds.

\section{The Messenger Threshold}
\label{sec:messenger}
We now show that a realistic messenger threshold at
$F_\varphi \sim 50\TeV$ arises naturally in simple anomaly-mediated models.
We include a UV divergent tadpole contribution
that was neglected in \Ref{HL}
that does not change the main conclusions.

The basic observation \cite{NW} is that
$F_\varphi \ne 0$ generates tree-level mass terms via
holomorphic \Kahler terms of the form
\beq[theKterms]
\De \scr{L} &= \myint d^4\th\, \frac{\varphi^\dagger}{\varphi} \,
c \, \bar{\Phi} \Phi\,
+ \hc
\\
\eql{theKterms2}
&= c \, F_\varphi^\dagger \myint d^2\th\, \varphi^{-1} \bar\Phi \Phi\, + \hc
\eeq
Here $c$ is a dimensionless constant
and $\bar\Phi$, $\Phi$ are vectorlike chiral fields.

Holomorphic \Kahler terms may not be familiar, but arise naturally from
integrating out heavy states at tree level.
For example,
consider the coupling of $\bar\Phi \Phi$ to heavy singlet
fields $\Si, \bar\Si$ via
\beq
\De W = \ka \varphi^2 \Si + M \varphi \bar\Si \Si
+ \sfrac 12 \la \Si \bar\Si^2
+ y \Si \bar\Phi \Phi .
\eeq
This has a $U(1)_R$ symmetry with
\beq[U1Rexample]
R(\Si) = 2,
\quad
R(\bar{\Si}) = R(\bar\Phi \Phi) = 0.
\eeq
Note that the $U(1)_R$ symmetry \Eq{U1Rexample} allows
the holomorphic \Kahler term but forbids a superpotential
mass term $\int d^2\th\, \bar\Phi \Phi$.
We now show that such a term is generated with an order-1
coefficient if $\ka \sim M^2$ and $\la, y \sim 1$.
The VEVs are
\beq
\avg{\Si} = 0,
\quad
\avg{\bar\Si} = \frac{-M \pm \sqrt{M^2 - 2\la \ka}}{\la}.
\eeq
The messengers are therefore massless at this order,
while $\Si, \bar\Si$ have masses of order $M$.
Integrating out $\Si, \bar\Si$ gives a vanishing superpotential
and a holomorphic \Kahler term of the form \Eq{theKterms} with
\beq
c = \left( -1 \pm \frac{M}{\sqrt{M^2 - 2 \la \ka}} \right)
\frac{y}{\la}.
\eeq
Here we assumed that all couplings are real,
and $M^2 - 2\la \ka > 0$.
We conclude that holomorphic \Kahler\ terms are generically
present when allowed by symmetries.

To understand the effect of holomorphic \Kahler terms
like \Eq{theKterms} on the SUSY spectrum,
note that the messenger threshold has $F/M = -F_\varphi$.
For comparison, a superpotential mass term 
$\int d^2\th\, M \varphi \, \bar\Phi \Phi$
with $F/M = +F_\varphi$ gives threshold corrections
that put the SUSY breaking terms on the anomaly mediated
renormalization group
trajectory in the low-energy effective theory.
(There are additional corrections of order
$F / M^2 = 1/c \sim 1$, but
in practice these corrections are small
unless $F/M^2 = 1$ to high accuracy; see the appendix.)
If $\bar\Phi$, $\Phi$ are messengers, the term \Eq{theKterms}
gives a threshold correction that affects the scalar masses
like a supersymmetric mass, since these are even in $F/M$.
Therefore the slepton masses are negative below the
messenger threshold.
The gaugino masses are odd in $F/M$ and receive a nontrivial
correction at threshold.
Running from the messenger scale to the weak scale can
give positive slepton masses, but this requires a large
number of messengers and results in a very fine-tuned model
\cite{NW}.

These problems can be easily cured if
the messenger threshold includes singlets in addition to messengers
\cite{HL}.
These singlets naturally get SUSY-breaking VEVs of order $F_\varphi$,
and couplings between the singlets and the messengers naturally
give large threshold corrections to scalar masses that can give
a natural spectrum.
A simple example consists of a singlet $X$, and doublet
and triplet messengers $\bar{D}$, $D$ and $\bar{T}$, $T$,
with all dimensionless couplings:
\beq
K &= -\frac{\varphi^\dagger}{\varphi}
\left( \sfrac 12 c_X X^2 
+ c_D \bar{D} D + c_T \bar{T} T \right),
\\
W &= \frac{\la_X}{3!} X^3
+ \la_D X \bar{D} D
+ \la_T X \bar{T} T.
\eeq
This theory is only technically natural, since a linear
superpotential term $\De W \sim \varphi^2 X$ is allowed
by all symmetries.%
\footnote{%
A superpotential term $\De W \sim \varphi X^2$ can be forbidden
by a discrete $R$ symmetry $X(\th) \mapsto -X(i\th)$.}
%

Note that there is a UV divergent 1-loop linear term in $X$:
\beq
\De\scr{L}_{\rm eff} &= \myint d^4\th\, \left[
-\frac{c \la F_\varphi}{16\pi^2} 
 (\phi^{-1})^\dagger
\ln \frac{\La |\varphi|}{\mu}
\, X  + \hc \right]
\\
&= \frac{c\la F_\varphi^2}{16\pi^2} 
\left( \ln\frac{\La}{\mu} - \sfrac 12 \right)
\myint d^2\th\, X + \hc
\nonumber\\
\eql{tadpoleX}
&\qquad
+ \frac{c\la F_\varphi^3}{32\pi^2} X + \hc,
\eeq 
where 
\beq
c\la \equiv c_X \la_X + c_D \la_D + c_T \la_T.
\eeq
The UV divergent $\int d^2\th$ term tells us that we 
have an uncalculable coupling
\beq
\De W = \ka_X X.
\eeq
Note that there is no $\varphi$ dependence in this term,
since it arises from the SUSY breaking from $F_\varphi \ne 0$
as described above.
The coupling $\ka_X$ has the renormalization group
equation
\beq
\frac{d\ka_X}{d\ln\mu} = -\frac{c\la F_\varphi^2}{16\pi^2}
- \sfrac 12 \ka_X \ga_X,
\eeq
where
$\ga_X = d\ln Z_X/d\ln\mu$
is the wavefunction renormalization of $X$.
If we run down from a large scale $\La$
(\eg\ the GUT scale) the large logs compensate for the loop
suppression, and we expect
\beq
\ka_X \sim -\frac{c\la F_\varphi^2}{16\pi^2}
\ln \frac{\La}{F_\varphi}
\sim -c \la F_\varphi^2.
\eeq
The model is therefore described by the effective superpotential
\beq
\bal
W_{\rm eff} &= -a_X F_\varphi^2 X
- F_\varphi \varphi^{-1}
\left(
\sfrac 12 c_X X^2 + c_D \bar{D}D + c_T \bar{T}T \right)
\\
&\qquad + \frac{\la_X}{3!} X^3
+ \la_D X \bar{D} D
+ \la_T X \bar{T} T.
\eal
\eeq
with $a_X \sim +c\la$.
We have neglected the finite tadpole in \Eq{tadpoleX},
since it is loop suppressed (and not log enhanced).
Including it will give a 1-loop correction to the
VEVs and masses of the particles.
All SUSY breaking is then in the mass terms, which 
are all of order $F_\varphi$.
The potential has a tadpole term for $X$
\beq
\bal
V &= a_X c_X X - \sfrac 12 (c_X - \la_X a_X) X^2 + \hc
\\
&\qquad
+ c_X^2 |X|^2 + \scr{O}(X^3).
\eal
\eeq
and therefore $\avg{X} \ne 0$.
For example, for $|a_X| \ll 1$ and $|c_X| > 1$ the VEVs are dominated
by the linear and quadratic terms in the potential and we obtain
a stable minimum with
\beq
\avg{X} = -\frac{a_X}{c_X - 1} F_\varphi,
\qquad
\avg{F_X} = \frac{(2c_X - 3) a_X^2}{2(c_X - 1)^2} F_\varphi^2.
\eeq
Exact expressions for the case $a_X = 0$ can be found in
\Ref{HL}.
The important point is that
the coupling of $X$ to the messengers gives
them a general value of $F/M \sim F_\varphi$.

In a grand unified theory it is
natural to impose boundary conditions
at the unification scale
\beq[unifybdy]
c_D(M_{\rm GUT}) = c_T(M_{\rm GUT}),
\qquad
\la_D(M_{\rm GUT}) = \la_T(M_{\rm GUT}).
\eeq
Since all couplings run only due to wavefunction renormalization,
we have
\beq
\frac{d}{d\ln\mu} \left( \frac{c_{D,T}}{\la_{D,T}} \right)
= \sfrac 12 \ga_X \frac{c_{D,T}}{\la_{D,T}},
\eeq
which implies that
\beq
\frac{c_D}{\la_D} = \frac{c_T}{\la_T} \equiv s
\eeq
at all scales.
In this case, the messenger threshold for doublets and singlets
are controlled by the same SUSY breaking parameter.
Writing the messenger masses as
\beq
W_{\rm eff} = M_D \bar{D} D + M_T \bar{T} T,
\eeq
we have
\beq
\frac{F}{M} = \frac{F_{M_T}}{M_T} = 
\frac{\displaystyle \avg{F_X} - s |F_\varphi|^2}
{\displaystyle \avg{X} + s F_\varphi^\dagger}.
\eeq
In this model, the SUSY breaking spectrum 
(excluding the $\mu$ and $B\mu$ terms; see below)
is parameterized by $F_\varphi$ and $F/M$.
The details are given in the appendix, including
a derivation of the results to all orders in 
$F / M^2$.
This is in principle important because 
$F / M^2 \sim 1$ (since all mass terms are of order $F_\varphi$),
but in practice the higher order corrections
are numerically negligible
unless $F/ M^2 = 1$ to high accuracy.

The model described above does not address the $\mu$ problem.
We will see that a realistic model of the $\mu$ term 
requires only a modest complication of the threshold
at the scale $F_\varphi$.

%

\section{The $\mu$-$B\mu$ Sector}
\label{sec:muterm}
Adding a supersymmetric $\mu$ term to the Lagrangian in AMSB
gives $B\mu = F_\varphi \mu$, which generates a Higgs VEV of order
$F_\varphi \sim 50\TeV$.
This means that the $\mu$ term (or better, the Higgsino mass term)
\emph{must} originate in some other way in AMSB to avoid
a phenomenological disaster.
We present a simple relaxation mechanism to suppress $B\mu$.
Our model requires a small coupling of order $10^{-2}$ to
suppress $\mu$ relative to the messenger scale, but with this
one small parameter $B\mu$ automatically has the correct size.

The model has two additional singlet superfields fields $S$ and $A$ with
superpotential
\beq[Wsimple]
\De W = -a F_\varphi^2 A + \sfrac 12 \la A S^2 + \ep S H_u H_d
\eeq
and \Kahler potential
\beq[Ksimple]
\De K = -\frac{\varphi^\dagger}{\varphi}
\left( \sfrac 12 c S^2 \right).
\eeq
The linear term in \Eq{Wsimple} is required due to the UV divergent
tadpole, as discussed
in the previous section,
and we expect $a \sim +c\la$.
A linear term in $S$ is forbidden by the $Z_2$ symmetry
\beq[Z2ordinary]
S \mapsto -S,
\qquad
H_u H_d \mapsto -H_u H_d,
\eeq
with all other fields even.

Note that $A$ appears in the 
tree-level scalar potential only via the $|F_S|^2$ term, so minimizing
the potential with respect to $A$ implies $\avg{F_S} = 0$.
This model therefore generates a tree-level $\mu$ term without a tree-level
$B\mu$ term.
This is a relaxation mechanism for the $B\mu$ term similar to
that of \Ref{muGMSB}.
The correct size of the $\mu$ term is obtained for $\ep \sim 10^{-2}$,
and a $B\mu$ term of order $\ep F_\varphi / 16\pi^2$ is generated
at one loop.
All components of the fields $A$ and $S$ have masses of order $F_\varphi$
(assuming that $\la, c \sim 1$), so the theory below the scale $F_\varphi$
is the MSSM.

The smallness of $\ep$ is perfectly natural since it is the only
coupling that violates the discrete symmetry $S \mapsto -S$,
with all other fields even.
On the other hand, the model taken as a whole is only technically natural,
since a superpotential term
$\De W \sim \varphi^2 A$
is allowed by all symmetries,
and would ruin the relaxation mechanism described above.%
\footnote{%
Superpotential terms of the form
$\De W \sim \varphi A^2 + \varphi S^2 + A^3$ can be forbidden
by a $U(1)_R$ symmetry.}
%

We now analyze the model in detail.
The tree-level potential is
\beq\bal
V &= \left| \la A S - c F_\varphi S + \ep H_u H_d \right|^2
+| \sfrac 12 \la S^2 - a F_\varphi^2 |^2
+ \ep^2 |S|^2 \left( | H_u|^2 + |H_d|^2 \right)
\\
&\qquad
- \sfrac 12 c F_\varphi^2 S^2 + \hc
\eal\eeq
We minimize with respect to $S$ and $A$,
expanding about $H_u, H_d = 0$.
This gives
\beq[thevacuumsimplemodel]
\avg{A} &= \frac{c F_\varphi}{\la},
\\
\avg{S} &= \pm \frac{\sqrt{2(c + a \la)}\, F_\varphi}{\la},
\eeq
assuming that $c + a \la > 0$.
Note that $\avg{F_S} = 0$ due to the $A$ minimization
condition, so we obtain
\beq
\mu = \ep \avg{S},
\qquad
B\mu = 0.
\eeq
The minimum spontaneously breaks the discrete symmetry \Eq{Z2ordinary},
so this theory has domain walls.
These can be eliminated by adding additional 
naturally small couplings (\eg\ higher-dimension operators)
that break this symmetry.
All scalar and fermion components of $S$ and $A$
get masses of order $F_\varphi$.
These masses are all equal to $\la \avg{S}$,
suggesting a hidden unbroken SUSY in the model.

The hidden SUSY can be made manifest by a superfield field
redefinition
\beq
A = A' - \frac{c F_\varphi}{\la \varphi}.
\eeq
We then have
\beq
\myint d^4\th\, A^\dagger A
= \myint d^2\th\, A'^\dagger A'
- \left( \myint d^2 \th\, \frac{c F_\varphi^2}{\la} A'
+ \hc \right)
+ \mbox{constant}.
\eeq
The effective superpotential is therefore
\beq[Wefffancy]
W_{\rm eff} = -\left( a + \frac{c}{\la} \right)
F_\varphi^2 A'
+ \sfrac 12 \la A' S^2 
+ \ep S H_u H_d.
\eeq
There is no conformal compensator in the linear term
because it arises from the couplings
\Eqs{Wsimple} and \eq{Ksimple},
so the Lagrangian preserves SUSY at tree level.
We see that the VEVs \Eq{thevacuumsimplemodel} are
simply the supersymmetric vacuum $\avg{F_S} = \avg{F_{A'}} = 0$.
Expanding about this vacuum 
$S = \avg{S} + S'$
gives a superpotential
\beq[Wefffancyexp]
W_{\rm eff} = \la \avg{S} A' S' + \sfrac 12 \la A' S'^2 + \cdots.
\eeq
This explains why all components
of $A$ and $S$ have a mass equal to $\la \avg{S}$.
The hidden supersymmetry of this model is rather
special.
The relaxation mechanism does not depend on this,
and more complicated models will not necessarily have
this structure.

We now discuss the $B\mu$ term.
As we have seen above, it vanishes at tree-level, but there
are nonvanishing 1-loop contributions.
From \Eq{Wefffancy} we see that the tree-level potential
is actually supersymmetric.
The only SUSY breaking comes from the finite linear term
arising from the finite part of the $A$ tadpole (see \Eq{tadpoleX}).
In terms of the shifted fields $A'$ and $S'$ we have
\beq
V = -\frac{c\la F_\varphi^3}{32\pi^2} (A' + \hc)
+ |\la \avg{S}|^2 \left( |A'|^2 + |S'|^2 \right)
+ \cdots.
\eeq
We therefore obtain
\beq
\avg{A'} = \frac{1}{64\pi^2} \frac{c\la}{c + a \la} F_\varphi.
\eeq
The $B\mu$ term is therefore
\beq
B\mu = \ep \la \avg{A'} \avg{S} + B_{\rm AMSB} \mu
= (\la \avg{A'} + B_{\rm AMSB}) \mu,
\eeq
where
\beq
B_{\rm AMSB} = \sfrac 12 \left( \ga_{H_u} + \ga_{H_d} \right) F_\varphi.
\eeq
Using
\beq
\ga_{H_u} = -\frac{N_c y_t^2}{8\pi^2} + \cdots
\eeq
where $N_c = 3$ is the number of colors, the contribution to $B$
from the $\mu$-$B\mu$ sector is
\beq
\frac{\De B}{B_{\rm AMSB}} \sim \frac{\la^2}{4 N_c y_t^2} \frac{1}{(1 + a\la/c)^2},
\eeq
so the anomaly-mediated contribution naturally dominates
due to a combination of the large top Yukawa coupling, color factors,
and order-1 factors.

\section{Phenomenology}

\subsection{Spectrum}
We now discuss the spectrum of the models under consideration.
We consider the minimal model, with parameters
\beq
F_\varphi,\ F,\ M, \mu,\ N_{\rm mess},
\eeq
where $F_\varphi$ and $F$ are the anomaly- and gauge-mediated
SUSY breaking parameters,
$M$ is the messenger scale,
and $N_{\rm mess}$
is the number of ${\bf 5} \oplus \bar{\bf 5}$ messengers.
We impose the unification condition \Eq{unifybdy} to relate the
doublet and triplet messengers.
This means that the spectrum is qualitatively similar to
gauge-mediated models, so we predict for example
$m_{\tilde{q}}/m_{\tilde{\ell}_R} \sim \al_3/\al_1$.
We can also relax this assumption and get spectra with
(for example) colored and uncolored superpartner masses
more closely degenerate \cite{HL}.
We will not discuss this possibility here.

\begin{figure}[t]
\begin{center}
\centerline{\includegraphics[scale=1.2]{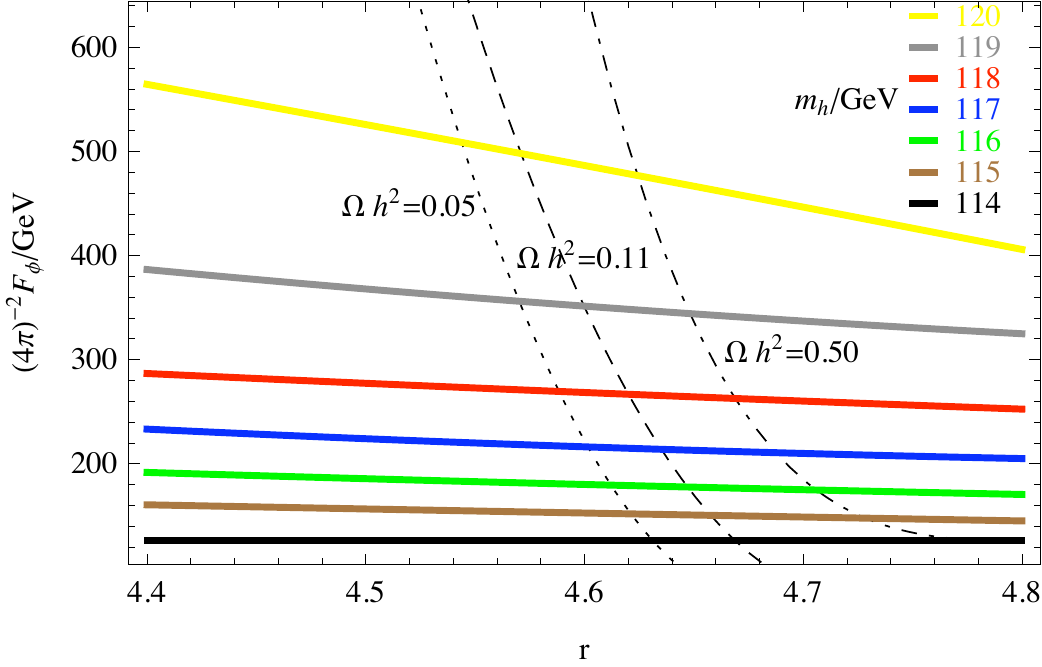}}
\begin{minipage}{5in}%
\caption{Lightest $CP$ even 
Higgs mass and neutralino LSP relic density in the minimal model
with $N_{\rm mess} = 1$.}
\end{minipage}
\end{center}
\end{figure}

\begin{figure}
\begin{center}
\centerline{\includegraphics[scale=1.2]{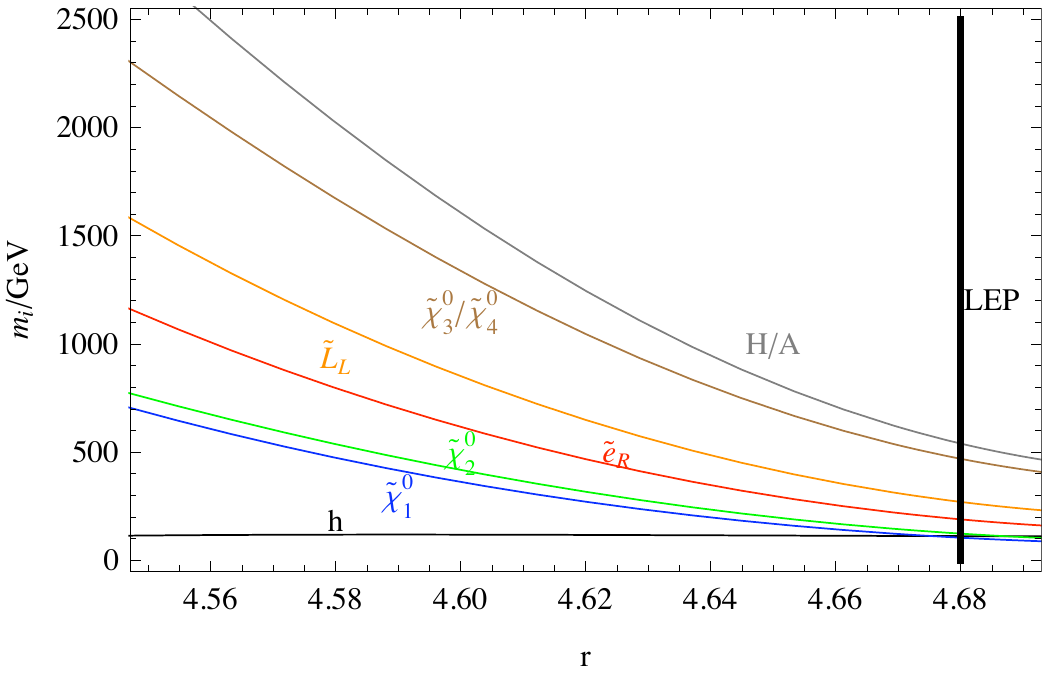}}
\begin{minipage}{5in}%
\caption{Uncolored superpartner masses in the minimal model
for parameters where the LSP has the
correct relic density.}
\end{minipage}
\end{center}
\begin{center}
\centerline{\includegraphics[scale=1.2]{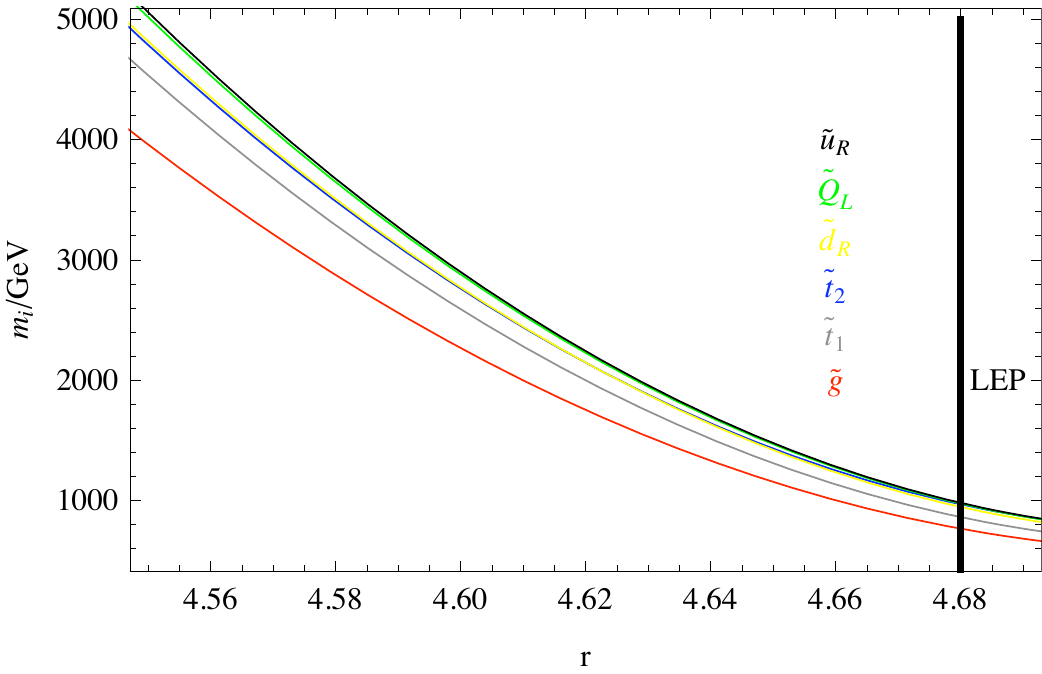}}
\begin{minipage}{5in}%
\caption{Colored superpartner masses in the minimal model
for parameters where the LSP has the
correct relic density.}
\end{minipage}
\end{center}
\end{figure}

The messenger scale $M$ is fixed to be of order $F_\varphi \sim 50\TeV$
in this model.
The SUSY breaking terms generated at the messenger threshold are
proportional to $F_\varphi$ and $F/M$ times a function of 
\beq
x = \frac{F}{M^2}.
\eeq
Positivity of messenger masses requires $x < 1$.
In practice, the dependence on $x$ is very weak unless
$x = 1$ to high accuracy (see appendix).
This dependence therefore drops out.
The SUSY breaking spectrum also depends on $M$ logarithmically
through the renormalization group, since $M$ determines the scale
at which we match the SUSY breaking masses onto the MSSM.
For example, changing the messenger scale by a factor of 2 changes the
spectrum at the level of a 1-loop correction, so this dependence 
can also be neglected.
The result of this is that the SUSY breaking spectrum essentially
depends only on the parameters
\beq
F_\varphi,\ 
\frac{F}{M},\ 
\mu,\ 
N_{\rm mess}.
\eeq

After we fix the Higgs VEV to its observed value, the theory
therefore depends on 2 continuous parameters, which we take to
be
\beq
M_{\rm SUSY} \equiv \frac{F_\varphi}{16\pi^2},
\qquad
r \equiv \frac{F/M}{F_\varphi}.
\eeq
For a range of positive $r$ we obtain a physical solution with
all scalar mass-squared terms positive, realistic electroweak
symmetry breaking, and a neutralino LSP.
In Fig.~1 we plot the Higgs mass and the LSP relic density for the
minimal model with $N_{\rm mess} = 1$.
We find that electroweak symmetry breaking occurs only for
$\mu > 0$.
These results were obtained using {\tt SuSpect} \cite{SUSPECT}.
The Higgs mass prediction is expected to be uncertain by 
about $\pm 2\GeV$.
We see that the Higgs mass is very close to the LEP bound,
and the correct relic density can be obtained.
The rest of the spectrum is shown in Figs.~2 and 3 for the models
along the line where $\Om h^2 = 0.11$.

\subsection{Dark Matter}
The spin-independent
direct detection dark matter-nucleon
cross section for these models
is plotted in Fig.~4.
These results we obtained using {\tt DarkSUSY} \cite{DarkSUSY}.
The cross sections are well below existing experimental
bounds.
Future ton-scale experiments
such as LUX/ZEPLIN \cite{LXZ}, SuperCDMS \cite{superCDMS},
and XENON1T \cite{XENON1T}
have an expected sensitivity to spin-independent
cross sections in the range
$10^{-46}$ to $10^{-48}~\mbox{cm}^2$
for these masses, and will therefore be able
to probe these models for WIMP
masses up to approximately 200~GeV.

\begin{figure}[t]
\begin{center}
\centerline{\includegraphics[scale=1.2]{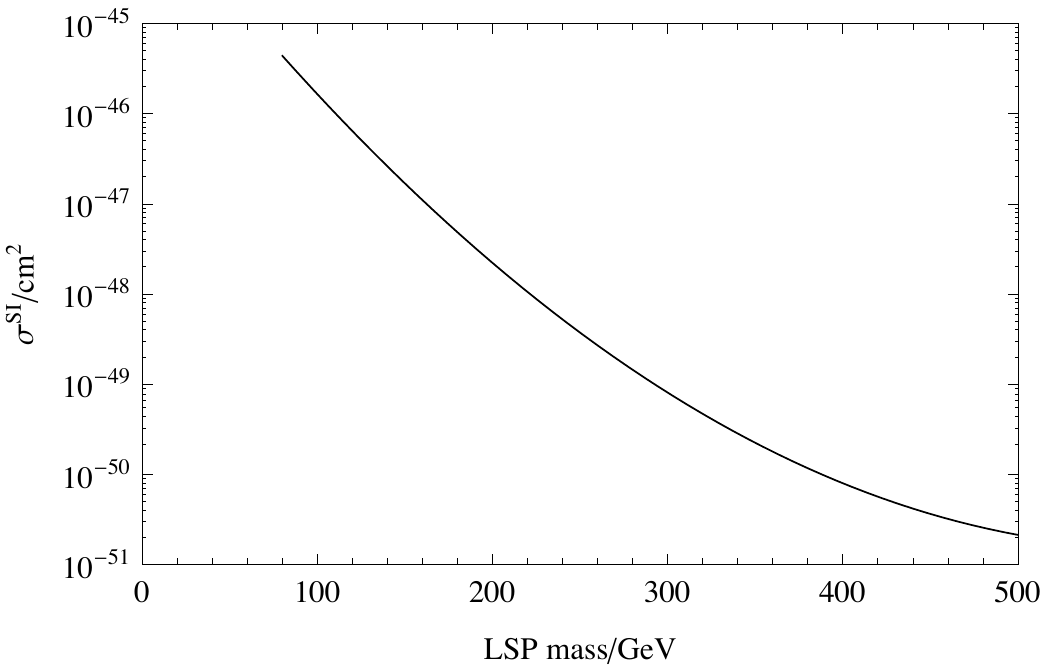}}
\begin{minipage}{5in}%
\caption{LSP direct detection cross section for
minimal model with correct relic abundance.}
\end{minipage}
\end{center}
\end{figure}

\subsection{Fine Tuning}
These models have the usual fine-tuning problem of SUSY models.
This arises because top and stop loops give a contribution
to the Higgs quadratic term 
\beq
\De m_{H_u}^2 \sim -\frac{N_c y_t^2}{4\pi^2} m_{\tilde{t}}^2
\ln \frac{M}{m_{\tilde{t}}}.
\eeq
The Higgs quadratic coupling must be of order the physical
Higgs mass $m_{h^0}^2$, so a rough measure of the fine-tuning is therefore
\beq
\mbox{tuning} \sim \frac{\De m_{H_u}^2}{m_{h^0}^2}.
\eeq
This is $\sim 100$ in our models.
We emphasize that this problem is shared by practically
all other SUSY models.
In particular, in models where the SUSY breaking terms 
are generated by standard model gauge interactions one always
has
\beq[stoprel]
\frac{m_{\tilde{t}}}{m_{\tilde{\ell}_R}} \sim \frac{\al_3}{\al_1} \sim 10,
\eeq
and therefore $m_{\tilde{t}} \sim \mbox{TeV}$.
This implies that the model is tuned
even without imposing the LEP bound on the Higgs mass.

There have been a number of solutions to the SUSY tuning problem
proposed in the literature.
One is exotic Higgs decays arising from additional structure
in the Higgs sector.
This may allow the Higgs to be light, but this will not
significantly reduce the fine-tuning in models satisfying \Eq{stoprel}.
Another possibility is new contributions to the Higgs
quartic coupling \cite{Higgsquartic}, which
reduce the fine tuning by allowing larger Higgs quadratic terms.
These generally affect only the Higgs sector, and the predictions 
for the squark, slepton, and gluino masses in this model still hold.
Another possibility is that tuning may be the result of anthropic
selection of a vacuum that
breaks electroweak symmetry \cite{SUSYtune}.

\section{Conclusions}
We have presented a minimal model of SUSY breaking that
naturally solves the SUSY flavor problem,
the $\mu$ problem, and
has a viable WIMP dark matter candidate.
SUSY is broken by anomaly mediation in the visible
sector, with a messenger threshold at the scale
$F_\varphi \sim 50\TeV$ generated by holomorphic
\Kahler terms.
Realistic $\mu$ and $B\mu$ terms are generated by a relaxation mechanism.
The models are similar to gauge mediated models in that standard
model gauge loops generate the SUSY breaking spectrum,
and therefore we have for example
$m_{\tilde q}/m_{\tilde\ell_R} \sim \al_3/\al_1$.
However the origin of SUSY breaking in anomaly mediation gives some
important differences with gauge-mediated models.
First, the gravitino is naturally at the scale $F_\varphi$
so the LSP is automatically a WIMP candidate.
Also, anomaly mediation gives rise to a $B$ term in the
Higgs potential from the top loop, and this generally
dominates the contribution from the sector that generates
the $\mu$ term.
This effectively eliminates one of the free parameters
in this model.

The minimal model depends on only 2 continuous parameters,
and the number of messengers.
We have shown that this model can give
realistic spectra with neutralino dark matter
that can be detected in upcoming experiments.
We believe these models deserve close study at the LHC.

\section*{Acknowledgements}
We thank N. Setzer and S. Spinner for pointing out an error
in the first draft.
This work was partially carried out at the Aspen Center for Physics.
This work was supported by DOE grant
DE-FG02-91-ER40674.

\appendix{Appendix A: SUSY Breaking}
In this appendix we derive the results for the SUSY breaking terms
at the scale $F_\varphi$.
The new feature is the consistent combination of anomaly and
gauge mediated contributions for the case $F / M \sim F_\varphi$
and $F / M^2 \sim 1$.

The SUSY breaking induced by the threshold at the scale $F_\varphi$
is best understood
by working in a formulation where all SUSY breaking is in the
higher components of the mass $M$ and UV cutoff $\La$.
For gaugino masses, we write the effective theory 
below the scale $M$ as
\beq[gauginomasseff]
\!\!\!\!\!\!\!\!\!
\scr{L}_{\rm eff}
= \myint d^2\th\, \tr(W^\al W_\al) 
\left[\tau(M, \La) + 
a_1 \frac{\bar{D}^2 M^\dagger}{M^2}
+ a_2 \left( \frac{\bar{D}^2 M^\dagger}{M^2} \right)^2
+ \cdots \right]
+ \hc,
\eeq
where $\tau$ is the gauge coupling superfield,
and the remaining terms are finite terms arising from 
integrating out the messengers at the scale $M$.
Here $\tau$, $M$, and $\La$ are all chiral superfields.
We have
\beq
\frac{\d}{\d\th^2} \frac{\bar{D}^2 M^\dagger}{M^2}
\propto \frac{F^2}{M^2},
\eeq
so that the terms proportional
to $a_0, a_1, \ldots$ are negligible for $F \gg M^2$.
However, we are interested in the case $F \sim M^2 \sim F_\varphi^2$,
so we must include them to all orders.
Terms such as $\bar{D}^2 \La^\dagger / \La^2$ may also arise from
integrating out physics at the cutoff,
but these give a vanishing contribution to SUSY breaking
in the limit $\La \to \infty$.
Mixed terms such as $\bar{D}^2 \La^\dagger / \La M$ cannot arise because
they violate decoupling of the scales $\La$ and $M$, namely all
terms that are unsuppressed as $\La \to \infty$
must be nonsingular as $M \to 0$.

The contribution from $\tau$ can be computed using the techniques
of \Refs{superspace}:
\beq
\frac{\d\tau}{\d\th^2} = \left( F_\varphi \frac{\d}{\d\ln\La}
+ \frac{F}{M} \frac{\d}{\d\ln M} \right) \tau
\eeq
We evaluate these derivatives at a scale $\mu < M$.
In this way we obtain the $\tau$ contribution to the
gaugino mass at the scale $M$:
\beq
m_{1/2}(M) = \frac{1}{g}
\left[ - F_\varphi \be'_g + \frac{F}{M} (\be'_g - \be_g) \right]
+ \cdots
\eeq
The limit $F / M \to F_\varphi$ corresponds to a superpotential
mass term, and gives
\beq
m_{1/2} \to \frac{F_\varphi}{g} \be_g + \cdots
\eeq
which is the anomaly-mediated contribution computed in the
effective theory below the scale $M$.
This is the famous decoupling of thresholds in anomaly
mediation.
This decoupling can be viewed as a consequence of the identity
\beq
\frac{\d}{\d\ln \La} + \frac{\d}{\d\ln M} + \frac{\d}{\d\ln\mu} = 0,
\eeq
where $\mu$ is the renormalization scale.
In the limit of an explicit superpotential mass term
$F / M \to F_\varphi$ this gives
\beq
\frac{\d}{\d\th^2} \to -F_\varphi \frac{\d}{\d\ln\mu},
\eeq
so the gaugino mass arising from $\tau$ is given by the
anomaly-mediated contribution in the effective theory in this limit. 

This decoupling is violated by the higher order
terms $a_1, a_2, \ldots$ in \Eq{gauginomasseff}.
These involve higher powers of
\beq
x =  \frac{F}{M^2}
\eeq
and correspond to computing the gauge-mediated contribution
to all orders in $x$.
This calculation was performed in \Refs{GMSBloop},
giving the result
\beq
m_{1/2}(M) = \frac{1}{g}
\left[ - F_\varphi \be'_g + \frac{F}{M} (\be'_g - \be_g) G(x) \right],
\eeq
where
\beq
\be_g = \frac{d g}{d\ln\mu}
\eeq
is the beta function for the gauge coupling $g$
in the effective theory below the scale $M$, and 
the primed quantities refer to the theory above the scale $M$.
The function $G$ is given by
\beq
G(x) = \frac{1}{x^2} \left[ (1 + x) \ln(1 + x) + (1 - x) \ln(1 - x) \right].
\eeq
For $x \to 0$, $G(x) \to 1$, while for 
$x \to 1$, $G(x) \to \ln 4 \simeq 1.4$.


The scalar masses are a bit more subtle because
there are in general
mixed gauge- and anomaly-mediated contributions.
The effective theory below the scale $M$ contains terms
\beq[scalarwavefcneff]
\scr{L}_{\rm eff} =
\myint d^4\th\, Q^\dagger Q \left[
Z(|M|, |\La|) + c_1 \left| \frac{D^2 M}{M^2} \right|^2
+ c_2 \left| \frac{D^2 M}{M^2} \right|^4
+ \cdots \right],
\eeq
where $Z$ is the wavefunction renormalization constant.
Here $Z$, $|\La| = (\La^\dagger \La)^{1/2}$,
and $|M| = (M^\dagger M)^{1/2}$ are real superfields.
Again, contributions involving \eg\ $D^2 \La / \La M$
are absent.
The contribution from $Z$ can be computed from
\beq
\frac{\d Z}{\d\th^2} = \frac{1}{2}
\left(F_\varphi \frac{\d}{\d\ln|\La|}
+ \frac{F}{M} \frac{\d}{\d\ln |M|} \right) Z,
\eeq
and we obtain for the scalar mass 
\beq
m_0^2(M) &= -\frac{\d}{\d \bar{\th}^2}
\frac{\d}{\d\th^2} \ln Z
\nonumber\\
&= -\frac{1}{4}
\biggl\{ F_\varphi^2 \frac{\d \ga'}{\d g'_i} \be'_i
- 2 F_\varphi \frac{F}{M}
\left(\frac{\d\ga'}{\d g'_i} - \frac{\d\ga}{d g_i} \right)
\be'_i
\nonumber\\
&\qquad\qquad
- \left( \frac{F}{M} \right)^2
\left[ \frac{\d\ga}{\d g_i}
\left( \be'_i - \be_i \right) 
- \left( \frac{\d\ga'}{d g'_i} - \frac{\d \ga}{\d g_i} \right)
\be'_i \right]
\biggr\}
+ \cdots
\eeq
where $g_i$ are the couplings in the theory and
$\ga =  d\ln Z / d\ln \mu$
is the anomalous dimension of the scalar field.
We can again check that for $F / M \to F_\varphi$
this reduces to the anomaly-mediated contribution
computed in the effective theory below the scale $M$.

The terms proportional to $c_1, c_2, \ldots$ 
in \Eq{scalarwavefcneff} give corrections that are higher order
in $x$, corresponding to the full gauge-mediated contribution.
Thus we obtain the full contribution
\beq
m_0^2(M) &= -\frac{1}{4}
\biggl\{ F_\varphi^2 \frac{\d \ga'}{\d g'_i} \be'_i
- 2 F_\varphi \frac{F}{M}
\left(\frac{\d\ga'}{\d g'_i} - \frac{\d\ga}{d g_i} \right)
\be'_i
\nonumber\\
&\qquad\qquad
- \left( \frac{F}{M} \right)^2
\left[ \frac{\d\ga}{\d g_i}
\left( \be'_i - \be_i \right) 
- \left( \frac{\d\ga'}{d g'_i} - \frac{\d \ga}{\d g_i} \right)
\be'_i \right] H(x)
\biggr\},
\eeq
where 
\beq
\!\!\!\!\!\!\!\!\!\!
H(x) = \frac{1 + x}{x^2}
\left[ \ln(1 + x) - 2\, \mbox{Li}_2\left( \frac{x}{1+x} \right)
+ \frac 12\, \mbox{Li}_2 \left( \frac{2x}{1 + x} \right)
\right] + ( x \leftrightarrow -x).
\eeq
The absence of corrections to the mixed terms is due to the fact 
that there are no corrections of the form $D^2 \La / \La M$ in
\Eq{scalarwavefcneff}.
For fields that do not have tree-level couplings to the messengers,
we have $\ga' = \ga$, and the 
mixed terms vanish.

\end{document}